# Aerosol Composition of Hot Giant Exoplanets Dominated by Silicates and Hydrocarbon Hazes

Peter Gao[1,*], Daniel P. Thorngren[2], Elspeth K. H. Lee[3], Jonathan J. Fortney[4], Caroline V. Morley[5], Hannah R. Wakeford[6], Diana K. Powell[4], Kevin B. Stevenson[7], and Xi Zhang[8]

[1]Department of Astronomy, University of California Berkeley, Berkeley, CA 94720, USA

[2]Institute for research on exoplanets, Universite 'de Montreal, Montreal, QC, H3C 3J7, Canada

[3]Department of Physics, University of Oxford, Oxford OX1 3PU, United Kingdom

[4]Department of Astronomy and Astrophysics, University of California Santa Cruz, Santa Cruz, CA 95064, USA

[5]Department of Astronomy, University of Texas at Austin, Austin, TX 78712, USA

[6]School of Physics, University of Bristol, HH Wills Physics Laboratory, Tyndall Avenue, Bristol BS8 1TL, UK

[7]Johns Hopkins Applied Physics Laboratory, Laurel, MD 20723

[8]Department of Earth and Planetary Sciences, University of California Santa Cruz, Santa Cruz, CA 95064, USA

*Corresponding author: gaopeter@berkeley.edu

**Aerosols are common in the atmospheres of exoplanets across a wide swath of temperatures, masses, and ages**[1-3]. **These aerosols strongly impact observations of transmitted, reflected, and emitted light from exoplanets, obfuscating our understanding of exoplanet thermal structure and composition**[4-6]. **Knowing the dominant aerosol composition would facilitate interpretations of exoplanet observations and theoretical understanding of their atmospheres. A variety of compositions have been proposed, including metal oxides and sulphides, iron, chromium, sulphur, and hydrocarbons**[7-11]. **However, the relative contributions of these species to exoplanet aerosol opacity is unknown. Here we show that the aerosol composition of giant exoplanets observed in transmission is dominated** **by silicates and hydrocarbons. By constraining an aerosol microphysics model with trends in giant exoplanet transmission spectra, we find that silicates dominate aerosol opacity above planetary equilibrium temperatures of 950 K due to low nucleation energy barriers and high elemental abundances, while hydrocarbon aerosols dominate below 950 K due to an increase in methane abundance. Our results are robust to variations in planet gravity and atmospheric metallicity within the range of most giant transiting exoplanets. We predict that spectral signatures of condensed silicates in the mid-infrared are most prominent for hot (>1600 K), low-gravity (< 10 m s$^{-2}$) objects.**

We simulate the vertical and size distribution of exoplanet aerosol particles using a 1-dimensional aerosol microphysics model. The aerosol particles are assumed to form via two possible paths: cloud formation through thermochemical reactions, and haze formation through methane photochemistry[8,10,12]. Cloud distributions are computed by explicitly calculating and balancing rates of particle nucleation, condensation, coagulation, evaporation, and transport via sedimentation and diffusion[13]. In particular, we consider the impact of nucleation energy barriers, which may inhibit formation of some proposed cloud species. Only coagulation and transport are considered for the computation of haze distributions, while the haze production rate is parameterized from comparing methane upwelling rates with methane photodissociation rates.

We generate aerosol distributions for a grid of 1-dimensional, globally averaged giant exoplanet model atmospheres with equilibrium temperatures $T_{eq}$ between 700 and 2800 K, surface gravities of 4, 10, and 25 m s$^{-2}$, and atmospheric metallicities of solar and 10 times solar. We compare synthetic transmission spectra computed from our model atmospheres to vertical aerosol distributions inferred from the amplitude of molecular features in exoplanet transmission spectra (Supplementary Figure 1). Recent compilations of the amplitude of the water absorption band near 1.4 µm[14,15] for warm giant exoplanets have revealed non-monotonic trends in aerosol distributions with $T_{eq}$: planets with $T_{eq}$ > 2300 K and 1100 K < $T_{eq}$ < 1600 K have increasing cloudiness with increasing temperature, while the opposite trend exists for planets with 1600 K < $T_{eq}$ < 2300 K and $T_{eq}$ < 1100 K[15]. By constraining our model with observations of a large sample of objects across an extensive parameter space, we minimize the degeneracies that typically arise when comparing exoplanet aerosol models to data from a single object or a smaller sample of objects[4,16,17].

We find that the observed trends cannot be explained by aerosol-free atmospheres but can be reproduced by our cloudy models (Figure 1). In particular, we find that the aerosol opacity in transmission is dominated by silicate clouds above ~950 K, with minor contributions from aluminum oxide and titanium dioxide clouds, and hydrocarbon hazes below 950 K (Figure 2a). While it is expected that silicate clouds are abundant, as Mg and Si are two of the most plentiful elements that can be incorporated into exoplanet clouds[9], clouds composed of elements of similar abundance like iron and metal sulphides (e.g. $Na_2S$) have much lower opacities in transmission. This is caused by the notably higher nucleation energy barriers of these condensates stemming from their high surface energies, which inhibits cloud formation (Figure 3; also see Methods and Supplementary Table 2). Thus, by explicitly computing the nucleation and condensation rates associated with each cloud composition, we are able to exclude most of the proposed cloud compositions from influencing aerosol opacities in near-infrared transmission. The formation of hydrocarbon hazes below ~950 K is also expected due to the transition of the primary carbon reservoir from carbon monoxide to methane[20], which acts as the source of hydrocarbon hazes for many objects in the Solar System[21,22]. However, because these hazes form at very low pressures, they mask the signature of condensate clouds and thus dominate near-infrared transmission opacity.

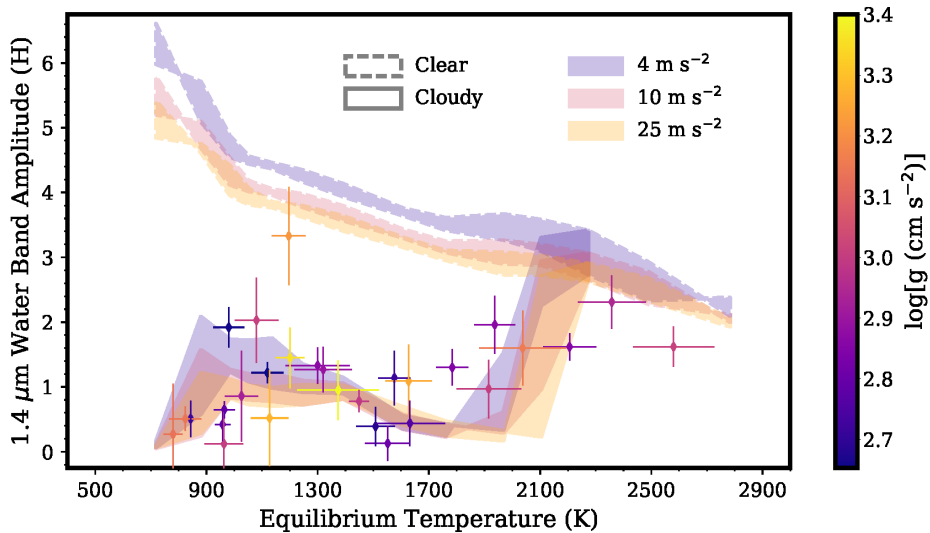

**Figure 1. Exoplanet cloudiness as a function of equilibrium temperature, gravity, and atmospheric metallicity.** Comparison between the observed[15] (points) and model (shaded regions) difference in planetary radii between the maximum in the 1.4 μm water band and the minimum in the adjacent J or H bands in units of the atmospheric scale height, as a function of planet equilibrium temperature and gravity. Models with gravities at 1 bar of 4 m s$^{-2}$ (purple), 10 m s$^{-2}$ (pink), and 25 m s$^{-2}$ (orange) are shown; the shaded region for each color indicates the spread between atmospheric metallicities of 1x and 10x solar. Shaded regions bordered by solid lines are for models with aerosols, while those bordered by dashed lines are for models without aerosols (clear atmospheres). The base-10 log of gravity at 1 bar of the observed planets are shown by the color of the points. The error bars on the observations represent 1 standard deviation uncertainties.

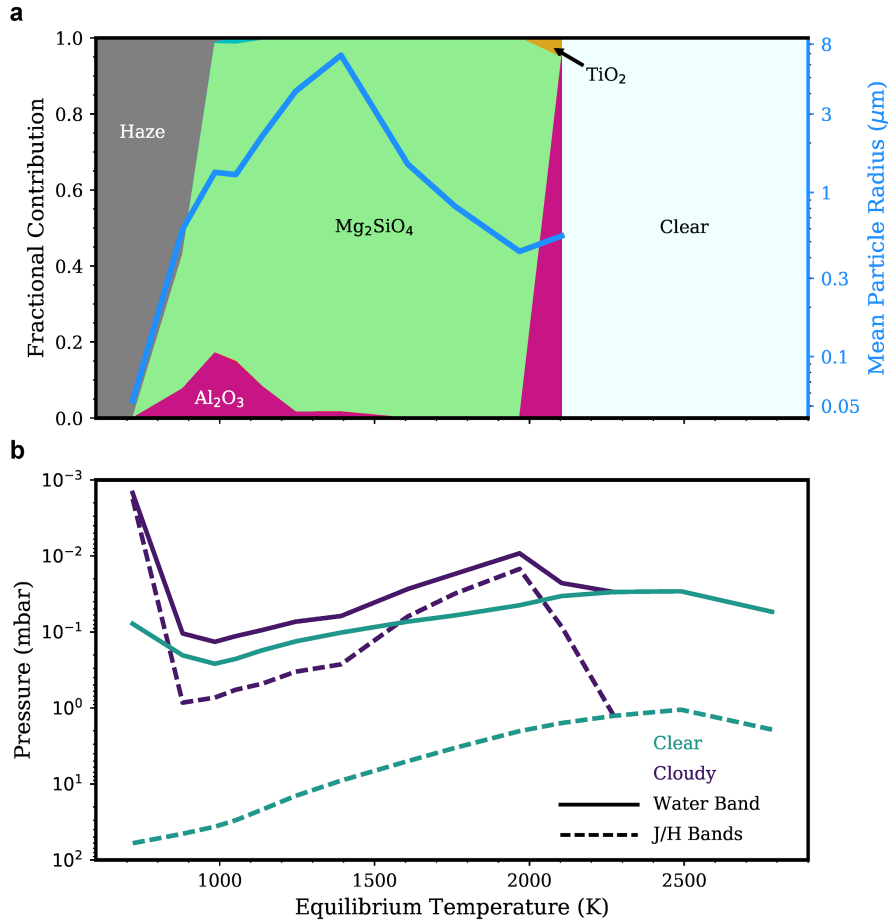

**Figure 2. Evolution of exoplanet aerosols with temperature. a,** Fractional contributions to the aerosol optical depth at the pressures probed in transmission at the wavelength that gives the minimum transit radii between 1.1 and 1.65 μm (in the J or H band), for a planet with 1 bar gravity of 10 m s$^{-2}$ and 10 x solar atmospheric metallicity. The cross section weighted mean aerosol particle radius at the same pressure is shown in blue. Note that the "clear" region only refers to the globally averaged sense; the permanent nightsides of warm giant exoplanets may be cool enough for cloud formation even when $T_{eq}$ > 2100 K[18]. **b,** The atmospheric pressure level probed by transmission spectra at the wavelength of minimum transit radii in the J or H bands (dashed) and at the wavelength of maximum transit radii in the 1.4 μm water band (solid), for cases with (purple) and without (turquoise) aerosols, for the same planets as in **a**. For clear planets, both the 1.4 μm water band and the J or H bands probe lower pressures with increasing temperature due to the temperature-dependence of the water absorption cross section[19]; this trend changes at $T_{eq}$ > 2500 K due to the thermal dissociation of water at low pressures, and at $T_{eq}$ < 950 K due to the increase in methane opacity at 1.4 μm. The existence of aerosols pushes the pressures probed in transmission to much lower values.

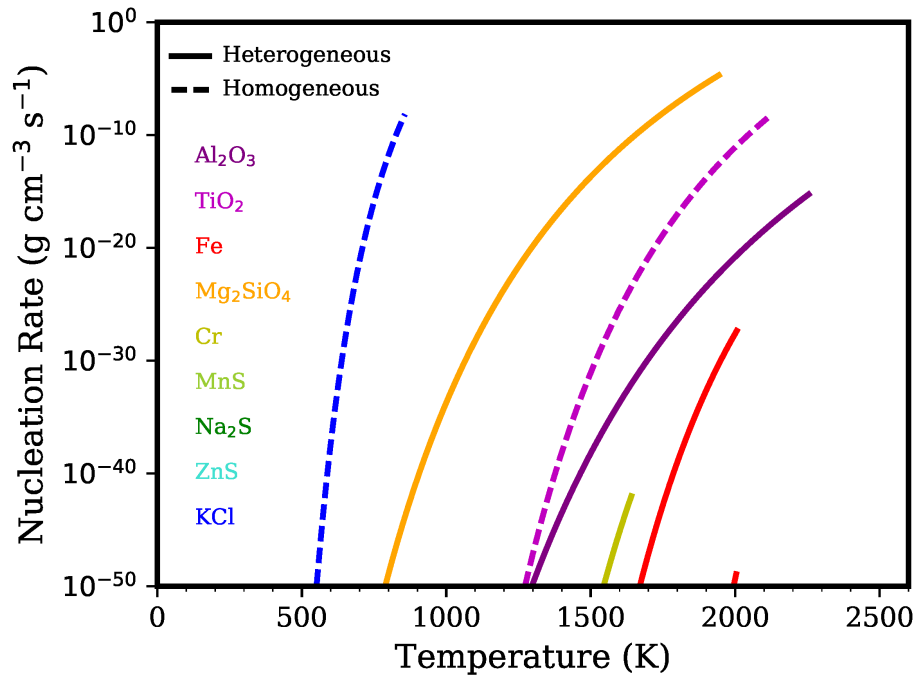

**Figure 3. Nucleation rates of exoplanet condensates.** Homogeneous (solid) and heterogeneous (dashed) nucleation rates of the condensates we include in our model, computed as a function of temperature for a fixed saturation ratio of 10 and a total atmospheric pressure of 10 bars. We assume an atmospheric metallicity of 10 times solar. We ignore solutions where the required saturation ratio implies a condensate vapor abundance greater than 10 times its solar abundance and where the temperature is greater than the condensation temperature. For the heterogeneous nucleation rate we assume a condensation nuclei radius of 0.5 microns and a number density of 10 cm$^{-3}$, consistent with values computed by our aerosol microphysics model. Nucleation rate curves for MnS, Na$_2$S, and ZnS are below the bottom boundary of the plot.

The observed trends in warm giant exoplanet cloudiness is a natural consequence of the dominance of only two types of aerosols (Figure 2b). The formation of silicate clouds at low pressures decreases the 1.4 µm water band amplitude with decreasing temperature for 1800 K < T$_{eq}$ < 2200 K. As temperature decreases further, the silicate cloud sinks into the atmosphere, reducing cloud opacity at low pressures; this increases the 1.4 µm water band amplitude with decreasing temperature for 950 K < T$_{eq}$ < 1800 K. Below 950 K, rising methane photodissociation rates lead to hydrocarbon haze formation at extremely low pressures (~1 µbar), resulting in the 1.4 µm water band amplitude decreasing with decreasing temperature.

Variations in warm giant exoplanet cloudiness with atmospheric metallicity and planet gravity are minor compared to that of equilibrium temperature, consistent with observations[15]. Changes in atmospheric metallicity between 1 and 10 x solar leads to the same magnitude changes in both water and condensate vapor abundance, resulting in much smaller changes in the 1.4 μm water band amplitude. Higher metallicities should greatly amplify these changes, however, due to the rapid decrease in the atmospheric scale height compared to our assumed scale height, computed assuming an atmospheric molecular weight of 2.3 g mol$^{-1}$, appropriate for solar metallicity. Likewise, the column atmospheric mass, and therefore the gas opacity, is inversely proportional to gravity, while the aerosol vertical transport time scale is also inversely proportional to gravity when transport is dominated by eddy diffusion, assuming the mixing length formulation for the eddy diffusion coefficient (see Methods), and the square of the gravity when transport is dominated by sedimentation. As gravity only varies by a factor of 5 in our sample, changes in gravity leads to only minute changes in the 1.4 μm water band amplitude, since both aerosol and gas opacity vary similarly.

The variations that are present when atmospheric metallicity and gravity are altered are dominated by changes in the temperature profile; higher metallicities and lower gravities lead to higher atmospheric temperatures for a given equilibrium temperature due to higher gas opacities caused by higher heavy element content and higher atmospheric mass, respectively. For cloudy atmospheres, this results in the largest differences between different atmospheric metallicity and gravity cases at equilibrium temperatures where cloud species first form at high altitudes, i.e. $T_{eq}$ ~ 2100 K, since the difference in pressures probed with and without clouds is maximized (see Figures 1 and 2b). Temperature and metallicity are also important in determining whether $CH_4$ or $CO$ is the dominant carbon species, leading to large differences between the different metallicity and gravity cases at $T_{eq}$ ~ 950 K, where photochemical hazes become the major aerosol opacity source.

Our results contrast with previous studies that predicted the importance of metal sulphide and chloride clouds[16,17], hydrocarbon hazes at $T_{eq}$ > 950 K[23], and sulfur hazes[11] for transmission, emission, and reflected light observations of exoplanets and brown dwarfs. By accounting for nucleation energy barriers, we found that the formation of metal sulphide clouds is strongly inhibited, and thus they do not contribute to aerosol opacity. In contrast, KCl cloud formation is highly efficient due to KCl's low nucleation energy barrier (Figure 3), but the KCl cloud layer is hidden beneath the hydrocarbon haze. As such, KCl clouds may contribute substantially to the aerosol opacity at $T_{eq}$ < 950 K for objects where haze formation is negligible, such as standalone brown dwarfs and directly imaged planets. High temperature hazes may contribute to warm giant exoplanet aerosol opacities, but their effect may only be important for $T_{eq}$ < 1300 K[23], while sulfur hazes derived from hydrogen sulfide form at temperatures lower than those considered here[11]. More generally, while the importance of silicate clouds and hydrocarbon hazes has been discussed in previous works[10,12], we are able to isolate them as the most important aerosol species here because we show that their evolution with equilibrium temperature and gravity is consistent with a diverse ensemble of observations, and that other aerosol species do not affect observations due to their high nucleation energy barriers.

A caveat of our study is that, by using 1-dimensional models we do not take into account the three-dimensionality of warm giant exoplanets, which are likely to be tidally locked to their host stars. This results in permanent daysides and nightsides with temperature differences increasing with increasing $T_{eq}$[24], leading to predicted[25] and observed[26] spatial inhomogeneity in the distribution of aerosols. These effects are unlikely to impact our conclusions substantially, however, as the terminator-averaged temperature profile observed in transmission should be more similar to the globally averaged profile that we use in our modeling than the more extreme day- and nightside profiles. In addition, because silicates and hazes dominate the aerosol opacity over wide ranges of temperatures, differences between the actual limb-average temperature profiles and our model profiles should not appreciably change the dominant aerosol species as a function of $T_{eq}$, except for minor variations at the transitions between clear atmospheres and different aerosol species. We also expect some differences in the vertical extent of the clouds due to the differences in temperature profiles. As the temperatures at the limbs are partially controlled by zonal winds that depend on the planetary rotation rate[27], variations thereof that are unaccounted-for in our modeling likely means that we have underestimated the scatter in the observed 1.4 μm water feature amplitude.

Even though we do not consider the day- and nightside cloud opacity of warm giant exoplanets explicitly in our modeling, our finding that only one type of cloud - silicates - dominates exoplanet cloud opacity over a wide range of temperatures has important implications for exoplanet emission and reflected light observations. For example, the brightness temperature of an atmosphere with an optically thick silicate cloud deck would be fixed to a value slightly below the condensation temperature of silicates where the cloud deck becomes optically thin, resulting in minimal variations in the atmospheric brightness temperature for 950 K < $T_{eq}$ < 2100 K. This is indeed what is observed for the nightsides of warm giant exoplanets, which all have brightness temperatures ~1100 K[28,29]. Meanwhile, the relatively high albedo of certain warm giant exoplanets like Kepler-7b[26] could also be explained by the dominance of silicate clouds, which are highly reflective at optical wavelengths[17]. We will rigorously test these hypotheses in a separate study that takes into account the different day- and nightside temperature profiles of warm giant exoplanets.

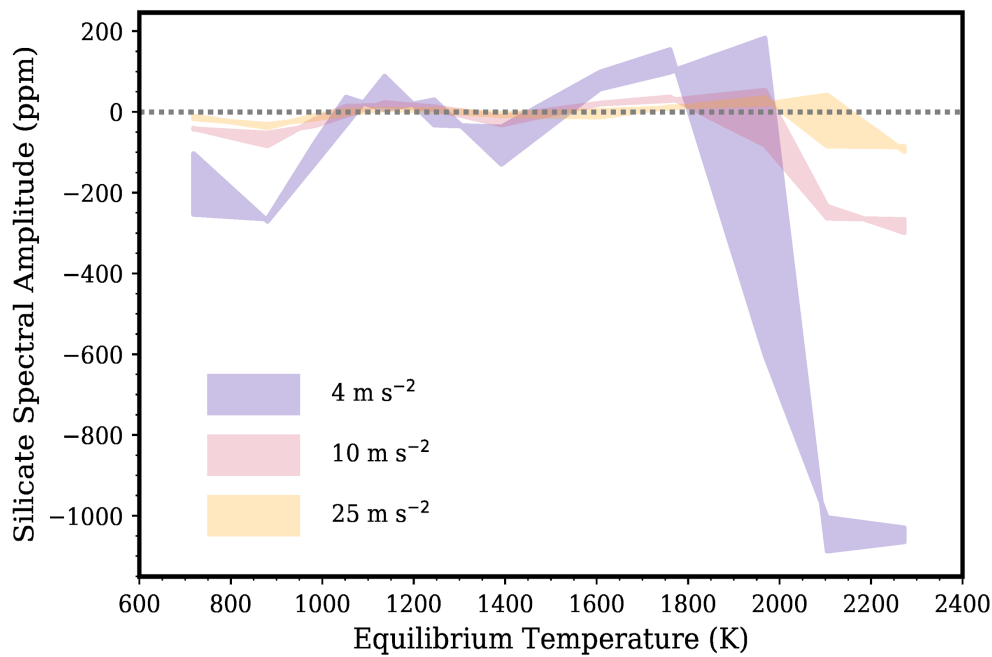

**Figure 4. Predictions for the amplitude of the 10 μm condensed silicate spectral feature in transmission.** The silicate feature amplitude as a function of planet equilibrium temperature for model planets with 1 bar gravity of 4 m s$^{-2}$ (purple), 10 m s$^{-2}$ (pink), and 25 m s$^{-2}$ (orange) with atmospheric metallicities between 1 x and 10 x solar (shaded regions). The horizontal dashed line marks zero amplitude, above which the feature is present.

Our prediction that silicates dominate aerosol opacity for most warm giant exoplanets can be tested by observing the condensed silicate spectral feature near 10 μm[30], which can be captured with the low-resolution spectroscopy mode of the Mid Infrared Instrument on the James Webb Space Telescope. The amplitude of the silicate feature is controlled by particle size and the height of the silicate cloud; the smaller the cloud particles and the higher the cloud is situated in the atmosphere, the more prominent the spectral feature. As the transmission spectrum is concave-up when the silicate feature is present and concave down otherwise, we use quadratic fits (Supplementary Figure 7) to search for its prominence. We predict that the silicate feature is most prominent for low gravity (< 10 m s$^{-2}$) planets with 1600 K < $T_{eq}$ < 2100 K, reaching above 100 ppm (Figure 4). As these planets host silicate clouds with small particles situated at low pressures (Figure 2), it is expected that they would exhibit high amplitude silicate spectral features. Evaluating the detectability of this feature requires conducting retrievals that take into account both the silicate opacity near 10 μm and the opacity of the gases that absorb at nearby wavelengths.

**Acknowledgements**

We thank M. S. Marley for his valuable insights and G. Fu for enlightening discussions. P. Gao acknowledges support from the NASA Postdoctoral Program and the 51 Pegasi b Fellowship from the Heising-Simons Foundation. E. K. H. Lee acknowledges support from the University of Oxford and CSH Bern through the Bernoulli Fellowship, and funding from the European community through the ERC advanced grant exocondense (# 740963). X. Zhang is supported by NASA Solar System Workings grant 80NSSC19K0791. H. R. Wakeford acknowledges support from the Giacconi Prize Fellowship at STScI, operated by AURA.

**Methods**

**Establishment of a trend in the data.** Transmission spectra obtained using the Hubble Space Telescope's Wide Field Camera 3 between 1.1 and 1.7 μm cover the J band (~1.22-1.3 μm), part of the H band (~1.48-1.8 μm), and a water absorption band (~1.36-1.44 μm). While the water band is sensitive to water vapor abundance, the J and H bands have comparatively lower molecular opacity, and therefore are sensitive to aerosol opacity. The vertical extent of aerosols can then be constrained by computing the difference in planetary radii, $A_H$, between the maximum in the water band and the minimum in the J or H bands in units of atmospheric scale height.

Several previous studies have computed $A_H$ or similar spectral indices for limited samples of exoplanets[3,14,15,31]. Here we use the $A_H$ values from ref. 15 but ignoring all exoplanets with observed radii below half a Jupiter radius, as they may possess metallicity higher than ~50 x solar. $A_H$ was computed[15] by taking the difference between the maximum and minimum planet radii between 1.3 and 1.65 microns of a model spectrum fit to the spectroscopic data[32], where fit optimization consisted of scaling the model spectrum and including a wavelength dependent slope. The resulting $A_H$ values are listed in Supplementary Table 1. However, we note that it is unclear if the minimum used in this wavelength range comes from the J- or H-band due to the wavelength dependent slope fit to the data. This may result in slight differences in the expected water feature amplitudes when comparing to $H_2O$-J or $H_2O$-H. Ref. 14 presents the $H_2O$-J index exclusively calculated on the data itself, rather than models fit to the data, though at the time only a dozen planets were available for study. We therefore use the index values presented in ref. 15 without any alteration aside from the radius cut described above as a comprehensive sample across a wide range of planet parameter space to test our models.

We check for nonlinear trends in $A_H$ as a function of planetary equilibrium temperature by fitting to the $A_H$ values increasingly higher order polynomials and calculating the Bayesian Information Criterion (BIC) for each fit,

(1) $BIC = \chi^2 + kLn(n)$

where k is the number of parameters, in this case the order of the polynomial plus 1, and n is the number of data points. If a robust nonlinear trend existed, then the BIC of fits with polynomials of order >1 should be smaller than that of order 1. We calculate the chi-square using the python package `scipy.odr`[33]. Supplementary Figure 2 shows the BIC as a function of the order of the polynomial fit to $A_H$, revealing that higher order polynomials do indeed fit the data better with minimal penalty due to increasing model complexity. Supplementary Figure 2 also shows the best fit polynomials with order 4 and greater, showing consistent non-monotonic behavior in $A_H$ as a function of equilibrium temperature. This assumes all systematics in the data[15] are accounted for within the uncertainties of the presented $A_H$ values, but we cannot account for the unknown selection of the J- or H-band as the minimum in the data fits.

**Construction of the background atmospheres.** We use an established thermal structure model for exoplanets and brown dwarfs[16,34-39] to generate our background atmospheres, which are described in detail in ref. 40. Briefly, we compute each model atmosphere's temperature-pressure (TP) and composition profiles assuming radiative-convective-thermochemical equilibrium, taking into account depletion of molecular species due to condensation and full heat redistribution (Supplementary Figure 3). The model planets were assumed to orbit sun-like stars, which is indeed the case for our sample of observed warm giant exoplanets (Supplementary Table 1). The atmospheres are divided into 60 layers, with the top layer at 1 μbar, and the bottom layer varying between 100 to 10000 bars. We also compute each planet's radius using a planetary interior model[41] to set the variation in gravity with altitude in the atmosphere. As the radius depends on the bulk metallicity, we set those of our model planets using the observed mass-metallicity relationship[41], though we fix the atmospheric metallicity to solar or 10 x solar, with the remaining heavy elements determining the core mass. We assume that our model planets have evolved to equilibrium, which is a valid assumption for most warm giant exoplanets[42], allowing us to set the intrinsic temperature using the empirical relationship from ref. 42.

We use eddy diffusion to parameterize large scale vertical mixing in the atmosphere. The strength of eddy diffusion depends on the mixing ratio gradient of the species and the eddy diffusion coefficient $K_{zz}$. The $K_{zz}$ value and profile are uncertain for exoplanet atmospheres, ranging across many orders of magnitude[43]. Here we use mixing length theory for the convective regions of the atmosphere[44], which has been successful in reproducing measured $K_{zz}$ values of giant planets in the Solar system and observations of exoplanet and brown dwarf spectra[45,46]. This results in $K_{zz}$ being proportional to the atmospheric scale height[44]. In radiative regions, we parameterize $K_{zz}$ with a minimum convective heat flux that falls off at one-third of the pressure scale height from the top of the convection zone, simulating convective overshoot (Supplementary Figure 3).

The atmospheric dynamic viscosity η and thermal conductivity κ are important for cloud formation when accounting for microphysics, as they control the sedimentation velocity of cloud particles and the rate with which latent heat released from condensation can be conducted away from the cloud particle, respectively. We use expressions of η and κ for $H_2$ from ref. 47

**Aerosol Microphysics Model.** We use the 1-dimensional Community Aerosol and Radiation Model for Atmospheres (CARMA) to simulate exoplanet aerosol distributions. CARMA computes the vertical and size distributions of aerosol particles by solving the discretized aerosol continuity equation, taking into account particle nucleation (homogeneous and heterogeneous), condensation, evaporation, and coagulation[48-51]. CARMA uses bins to resolve the particle size distribution, allowing for multiple particle modes to be simulated simultaneously and avoids the need to parameterize the size distribution using an analytical function, which may introduce errors in cloud optical properties of ~20-50%[25,52]. We refer the reader to our previous work[52] for a complete description of CARMA.

**Condensation sequence in exoplanet atmospheres.** We consider the formation of clouds composed of KCl, ZnS, $Na_2S$, MnS, Cr, $Mg_2SiO_4$, Fe, $TiO_2$, and $Al_2O_3$. These compositions are predicted by equilibrium condensation and kinetic cloud models[7-9] to be the most abundant condensates across our model temperature range.

The formation processes of our chosen cloud compositions are different for different species. KCl, Cr, Fe, and $TiO_2$ can undergo direct phase changes like water since the condensate molecule can exist as vapor in large abundance; these species may nucleate homogeneously or heterogeneously. ZnS, $Na_2S$, MnS, $Mg_2SiO_4$, and $Al_2O_3$, in contrast, form via thermochemical reactions. For example, ZnS does not exist as a molecule in the gas phase, and is thought to condense when Zn vapor reacts with $H_2S$ gas, forming ZnS condensate and $H_2$[10]. Thus, these clouds are more likely to form through surface reactions akin to heterogeneous nucleation, which requires the presence of cloud condensation nuclei (CCN). In addition, while equilibrium condensation predicts $CaTiO_3$ and other calcium-titanates as the major Ti condensate[53], kinetic models[8] predict that $TiO_2$ should be more prevalent due to the necessity of three-body reactions to form calcium-titanates, which are kinetically prohibitive.

We consider the homogeneous nucleation of seed particles/CCN, upon which other condensates can nucleate heterogeneously. We assume that nucleation converts the CCN into a cloud core that is completely enveloped by a shell of the condensing material, preventing any interaction between the core and the atmosphere. This is a simplified version of grain chemistry models[9] that allow for mixed particles where multiple species can simultaneously interact with the atmosphere. We restrict the maximum complexity of cloud particles to one shell overlying one cloud core (i.e. two compositions maximum per particle). Note that we do not explicitly treat surface reactions; instead we assume that the more abundant reactant is already present near or on the CCN and that the reaction occurs instantaneously, such that the nucleation rate is limited by the diffusion rate of the less abundant reactant to the CCN. To ensure mass conservation, when nucleation occurs the core mass fraction is stored in the model and reconstituted as the CCN particle when the shell evaporates.

We assume that the composition of the CCN is determined by which materials homogeneously nucleate the fastest and are relatively abundant, yielding $TiO_2$ and $KCl$[54]. $TiO_2$ nucleates at high temperatures (~2000 K), allowing it to act as CCN for $Al_2O_3$, Fe, $Mg_2SiO_4$, Cr, MnS, and $Na_2S$. KCl acts as CCN for ZnS, as they nucleate at lower temperatures, when most of the $TiO_2$ cloud mass would be at higher pressure levels. $TiO_2$ and KCl are treated as cloud particles first and foremost, which means they can grow by condensation and shrink by evaporation. They become cloud cores when the rate of heterogeneous nucleation of the nucleating material overtakes their growth rates[47].

Our explicit treatment of homogeneous and heterogeneous nucleation allows us to evaluate the importance of nucleation energy barriers, which may prevent cloud formation[52]. The nucleation energy is a function of the local (reaction) supersaturation[55] and material properties of the condensate, including its surface energy, molecular weight, and mass density (Supplementary Table 3)[52]. Rates of heterogeneous nucleation also depend on the contact angle of the condensate over the CCN; higher values of the contact angle lead to lower nucleation rates, and vice versa. The value of the contact angle $\theta_c$ is determined by the surface energies of the condensate and CCN, $\sigma_x$ and $\sigma_C$, respectively, and the interfacial energy between them $\sigma_{xC}$, which can be expressed via Young's relation for an ideal surface[56],

(2) $cos\theta_c = \frac{\sigma_C - \sigma_{xC}}{\sigma_x}$

For most of our considered condensates (Supplementary Table 2), $\sigma_x > \sigma_C$, while $\sigma_{xC}$ is unknown. If $\sigma_{xC} = 0$, then $cos\theta_c$ is just the ratio of the surface energies of the CCN and the condensate, yielding $\theta_c$ between 0° and 90°; as $\sigma_{xC}$ increases, so does $\theta_c$ until it equals $\sigma_x + \sigma_C$, at which point the contact angle becomes the maximum 180° and further increases in $\sigma_{xC}$ would no longer give a valid solution (i.e. no nucleation occurs). In this work we will assume that $\sigma_{xC} = 0$, which means we may overestimate the aerosol mass loading. In the case of $Mg_2SiO_4$, $\sigma_x < \sigma_C$, and so we assume a minimum contact angle of 0.1° to avoid numerical instability.

In addition to the contact angle, the desorption energy of a condensate molecule on the surface of the CCN is also important in determining the heterogeneous nucleation rate. Higher desorption energies >1 eV correspond to the formation of chemical bonds, while lower desorption energies ~0.1 eV are more indicative of van der Waals interactions[57-59]. As the desorption energies between our condensates and CCNs are not known, we choose an approximate midway value of 0.5 eV.

We assume a zero flux upper boundary and fix the condensed mass to zero at the lower boundary, as the temperatures there are always high enough to prevent condensation of all considered species. We compute the intersection of the saturation vapor pressure and the thermochemical equilibrium partial pressure of the limiting condensate vapor species (Supplementary Table 3) using GGchem[60], and set the lower boundary condensate vapor mixing ratio to the mixing ratio value at the intersection. In other words, we ignore any variations in the vapor profile below the cloud base due to equilibrium chemistry. For most of the considered condensates, this mixing ratio value corresponds to the elemental abundance of the limiting species (Supplementary Table 4). The only species where this is not true are KCl and $TiO_2$, the abundances of which are lower than the limiting elemental abundances (K and Ti, respectively) at the pressure and temperature of condensation, as shown in Supplementary Figure 4. This is due to competing species such as TiO for $TiO_2$ and atomic K and KOH for KCl.

We use 65 cloud particle mass bins in the model, each corresponding to masses twice that of the previous bin. The radius associated with the smallest mass bin is 1 Å for $TiO_2$, KCl, and homogeneously nucleated Fe and Cr; $2^{1/3}$ Å for $Al_2O_3$, $Mg_2SiO_4$, MnS, ZnS, and heterogeneously nucleated Fe and Cr; and $2^{2/3}$ Å for $Na_2S$. The bins are staggered in this fashion to take into account the radius bin mapping scheme of CARMA and maintain mass conservation. In CARMA, CCNs in a certain radius bin are mapped to cores of the same or a larger radius bin of the nucleating species. However, if the largest radius bin of the nucleating species has lower mass than the largest radius bin of the CCN due to the nucleating species having lower mass density, then the largest CCNs cannot be mapped and thus cannot act as CCNs, as they would be nucleating into a radius bin of lower mass, resulting in a loss of mass from the system. By staggering the bins, we ensure that the target bin always represents a more massive particle than the originating bin.

The model is initialized with our background atmosphere devoid of cloud particles, and condensate vapor only present at the bottom of the model atmosphere at their fixed mixing ratios. As the model marches forward in time, all condensate vapor are mixed upwards until they are either fully mixed in the atmosphere or they become supersaturated, at which point nucleation may occur. If $TiO_2$ or KCl are able to homogeneously nucleate, then all other condensates capable of nucleating heterogeneously on them that possess sufficiently high supersaturation would also nucleate. The particles are then transported via sedimentation and eddy diffusion until steady state is reached. We also include coagulation in our simulations, but only between particles with the same composition. We assume that the resulting particle is spherical like the particles that originally coagulated.

**Hydrocarbon hazes.** We model the production, sedimentation, diffusion, and coagulation of spherical haze particles. As an analog to cold, reducing atmospheres in the Solar System, we assume that the haze stems from methane photolysis and subsequent polymerization of photolysis products[21].

The physical and chemical processes that define the haze production rate are complex[11,21,61-62]. Here, we consider two endmember processes that limit haze production, the diffusion limit and the photon limit. In the diffusion limit, haze production depends on how fast the parent molecule can be replenished at the pressure level of production. We define the diffusion-limited haze production rate $P_{phot}$ as[63],

(3) $P_{phot} = N_{atm} K_{zz} \frac{df_{CH4}}{dz}$

where $N_{atm}$ is the atmospheric density, $K_{zz}$ is the eddy diffusion coefficient, and $df_{CH4}/dz$ is the methane mixing ratio gradient, which we approximate as $f_{CH4}/H_a$, or the complete loss of methane over a scale height $H_a$. We set the pressure of haze formation to 1 μbar, consistent with previous exoplanet photochemical and haze formation models[23,61], and use the values of $N_{atm}$, $K_{zz}$, $f_{CH4}$, and $H_a$ at that pressure level.

In the photon limit, haze production depends on the flux of stellar radiation capable of photolysis of the parent molecule. We define the photon-limited haze production rate $P_{Ly\alpha}$ (assuming Lyα dominates the photon flux) as

$$(4) \quad P_{Ly\alpha} = \frac{I_{Ly\alpha}}{4a^2} \frac{C_{CH4} f_{CH4}}{C_{H2O} f_{H2O} + C_{CH4} f_{CH4}}$$

where $I_{Ly\alpha}$ = 3.7 x $10^{11}$ cm$^{-2}$ s$^{-1}$ is the solar Lyα flux at 1 AU[64], a is the semi-major axes of the planet's orbit in AU, and $C_x$ and $f_x$ are the Lyα cross section and mixing ratio of molecule x, respectively, at the pressure level where photolysis occurs (1 μbar). This expression takes into account shielding due to photolysis of other species. For a solar composition atmosphere, by far the most abundant molecules that are readily photolyzed at Lyα are methane and water, where $C_{CH4}$ = 1.8 x $10^{-17}$ cm$^2$ and $C_{H2O}$ = 1.53 x $10^{-17}$ cm$^2$ (ref. 65), and so we only include these two molecules in our analysis. Other abundant molecules, such as $H_2$ and CO, do not have appreciable Lyα cross sections[65].

Supplementary Figure 5 shows the ratio of $P_{dif}$ to $P_{Ly\alpha}$ and reveals that the diffusion-limited production rate is always lower than the photon-limited production rate for the parameter space we have considered. Thus, we use the diffusion-limited production rate in our simulations.

We assume hydrocarbon soots as the haze composition[11], with a mass density of 1 g cm$^{-3}$ and a minimum particle radius of 50 nm, though for spherical particles the minimum particle radius has little impact on the haze optical depth at equilibrium[63]. We also assume that the haze does not interact with the condensate clouds.

**Transmission spectra.** We compute the extinction efficiency, single scattering albedo, and asymmetry factor of our simulated aerosol particles using the `pymiecoated` tool[66], which treats Mie scattering for layered spheres, taking into account contributions from both the core and the shell. The real and imaginary refractive indices of our considered aerosol species are shown in Supplementary Figure 6. We generate clear and cloudy transmission spectra following refs. 67,68 and include a correction for forward scattering by aerosol particles[69].

To compare model and observed $A_H$ values, we bin down the model spectrum to the resolution of the observations[32] and then divide by the best fit linear slope between 1.1 and 1.65 μm. We next calculate the difference between the maximum and minimum model planetary radii in the wavelength range and divide this value by the scale height, defined by the planet's equilibrium temperature and gravity at the 1 bar pressure level, to obtain the model $A_H$. We assume a mean molecular weight of the atmosphere of 2.3 g mol$^{-1}$ regardless of whether the atmospheric metallicity is solar or 10 x solar to be consistent with the scale heights assumed for the data[15].

We compute the amplitude of the 10 μm condensed silicate feature in our model spectra by fitting a second order polynomial to the wavelength region between 8.8 and 11.1 μm, then finding the amplitude of the polynomial within this wavelength range (Supplementary Figure 7).

**Methods References**

**Supplementary Information**

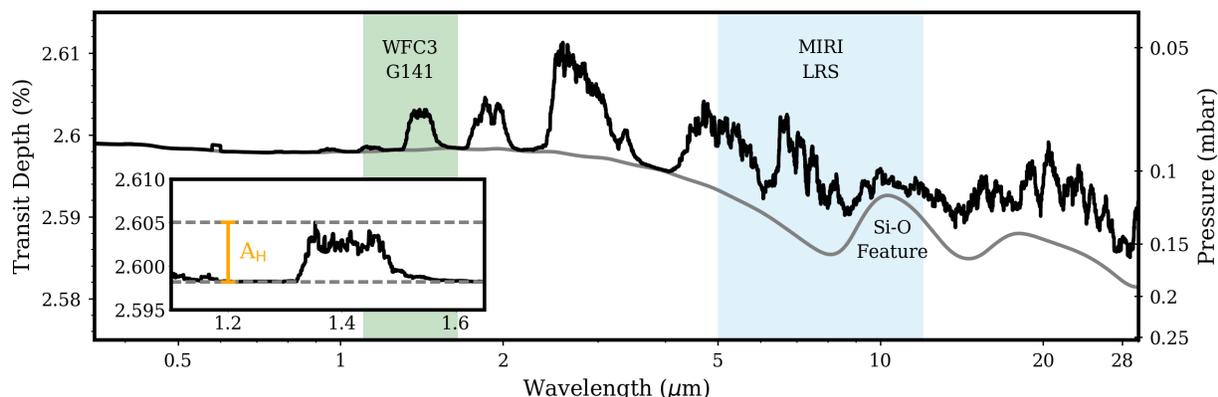

**Supplementary Figure 1. Probing cloudiness with transmission spectra.** Model cloudy transmission spectrum (black) for a hypothetical hot Jupiter orbiting a sun-like star at 0.025 AU with 1 bar gravity of 4 m s$^{-2}$ and a solar metallicity atmosphere. The spectrum is shaped by molecular absorption and cloud opacity, the latter shown in gray. The spectral regions considered in this work are shaded and the 10 µm silicate spectral feature is labeled. Inset: zoom in on the WFC3 G141 band, where the observed $A_H$ is defined as the difference between the gray dashed lines in units of atmospheric scale height.

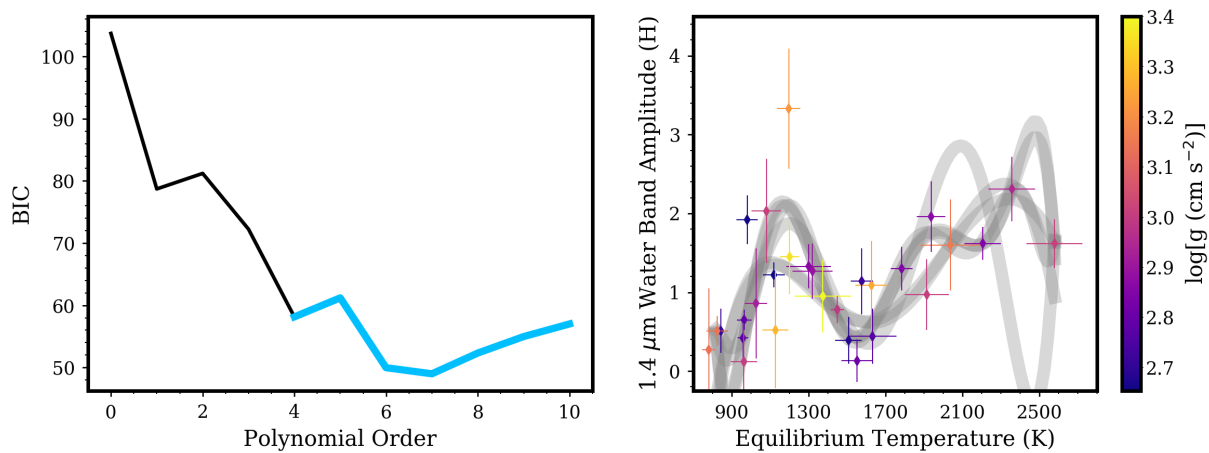

**Supplementary Figure 2. Analysis of trends in the data.** (Left) The Bayesian Information Criterion (BIC) as a function of the order of polynomials fit to the observations[1]. (Right) Polynomial fits of order 4 and above (i.e. the blue portion of the curve on the left). The oscillations for $T_{eq} > 2100$ K are due to the low number of data points there. The errorbars on the observations represent 1 standard deviation uncertainties.

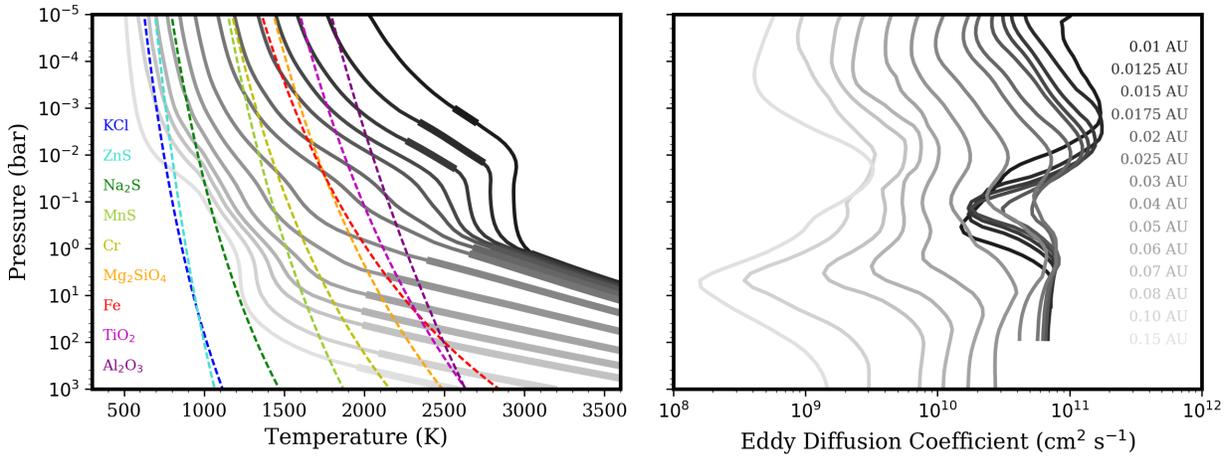

**Supplementary Figure 3. Model atmospheres.** Temperature-pressure (left) and eddy diffusion coefficient (right) profiles generated by our thermal structure model for planets with 10 x solar atmospheric metallicity and gravity at 1 bar of 10 m s$^{-2}$ placed at the labeled semi-major axes from a sun-like host star, compared to condensation curves of cloud species considered in this work. Convection zones are marked by thicker curves in the left figure.

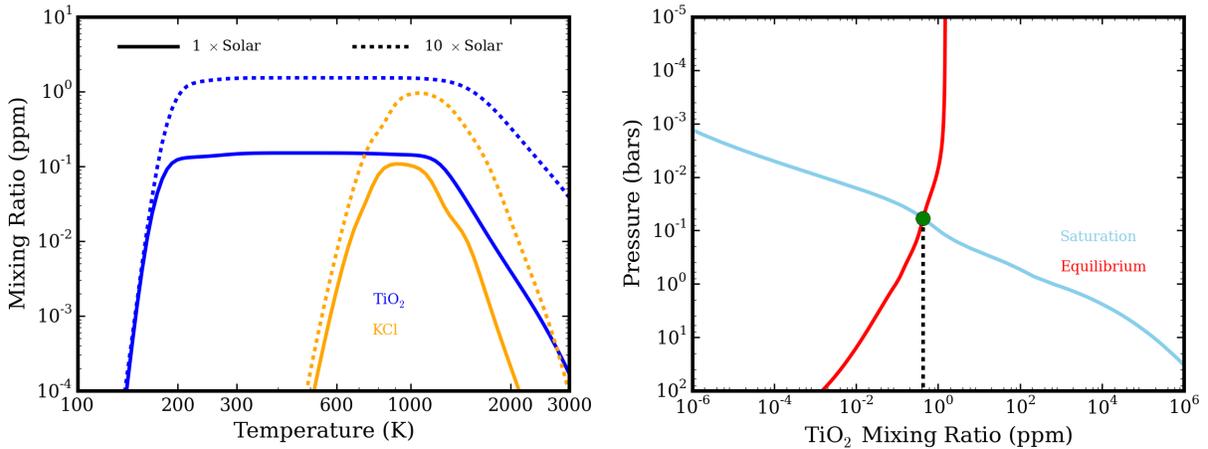

**Supplementary Figure 4. Combining equilibrium chemistry with condensation.** (Left) Mixing ratio of TiO$_2$ (blue) and KCl (orange) as a function of temperature assuming thermochemical equilibrium for solar metallicity (solid) and 10 x solar metallicity (dashed). (Right) We assume that the condensate vapor mixing ratio below the cloud base is constant and set by the mixing ratio where the equilibrium partial pressure equals the saturation vapor pressure in the atmosphere, despite the former varying with temperature, and thus atmospheric pressure level.

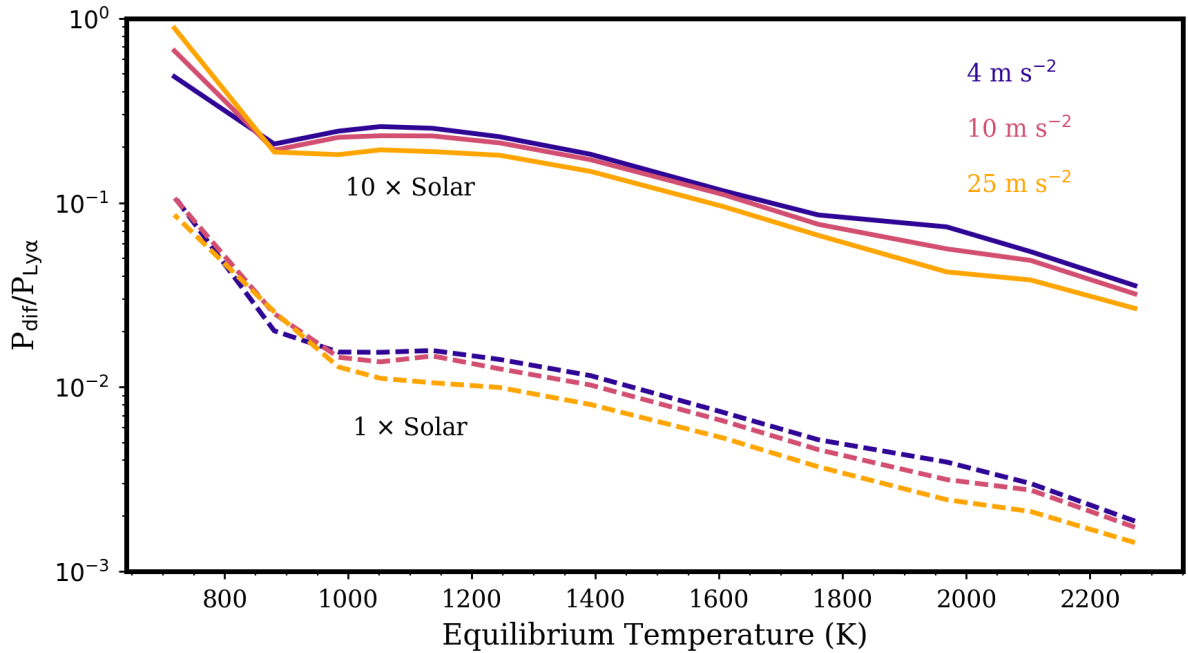

**Supplementary Figure 5. Haze production limits.** The ratio of methane diffusion-limited to Lyα-limited haze production rate as a function of equilibrium temperature for planets with gravities at 1 bar of 4 m s$^{-2}$ (purple), 10 m s$^{-2}$ (pink), and 25 m s$^{-2}$ (orange) and atmospheric metallicities of 1 x (dashed) and 10 x solar (solid). Note that all values sit below 1, indicating that all haze production in the parameter space explored here is limited by methane diffusion.

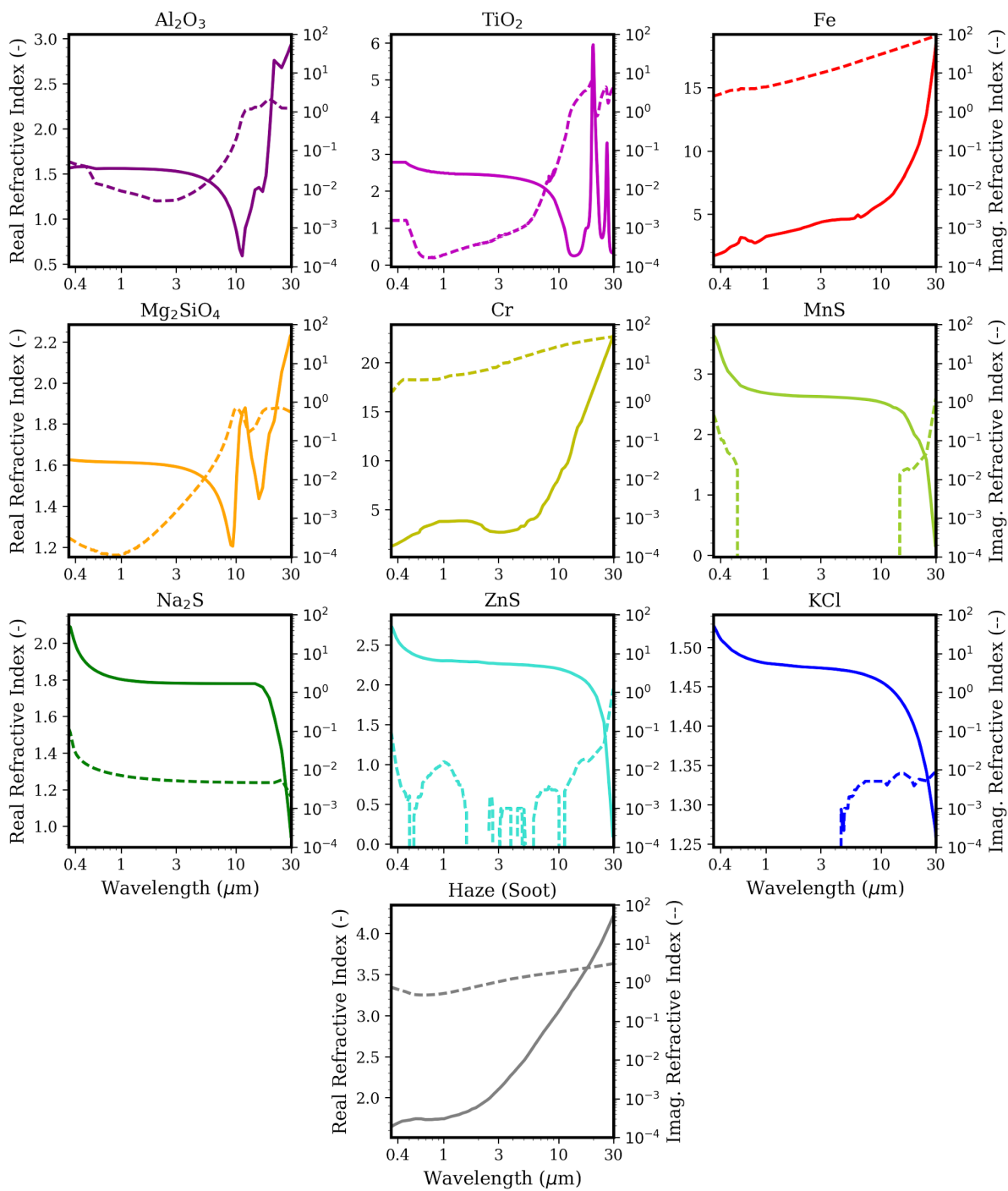

**Supplementary Figure 6. Refractive indices.** The real (left, solid) and imaginary (right, dashed) refractive indices of our considered condensates. Those of KCl, ZnS, Na$_2$S, MnS, and Cr are taken from ref. 2; those of TiO$_2$ are from refs. 3,4; those of Fe, Mg$_2$SiO$_4$, and Al$_2$O$_3$ are from ref. 5; and those of the hydrocarbon haze are from ref. 6.

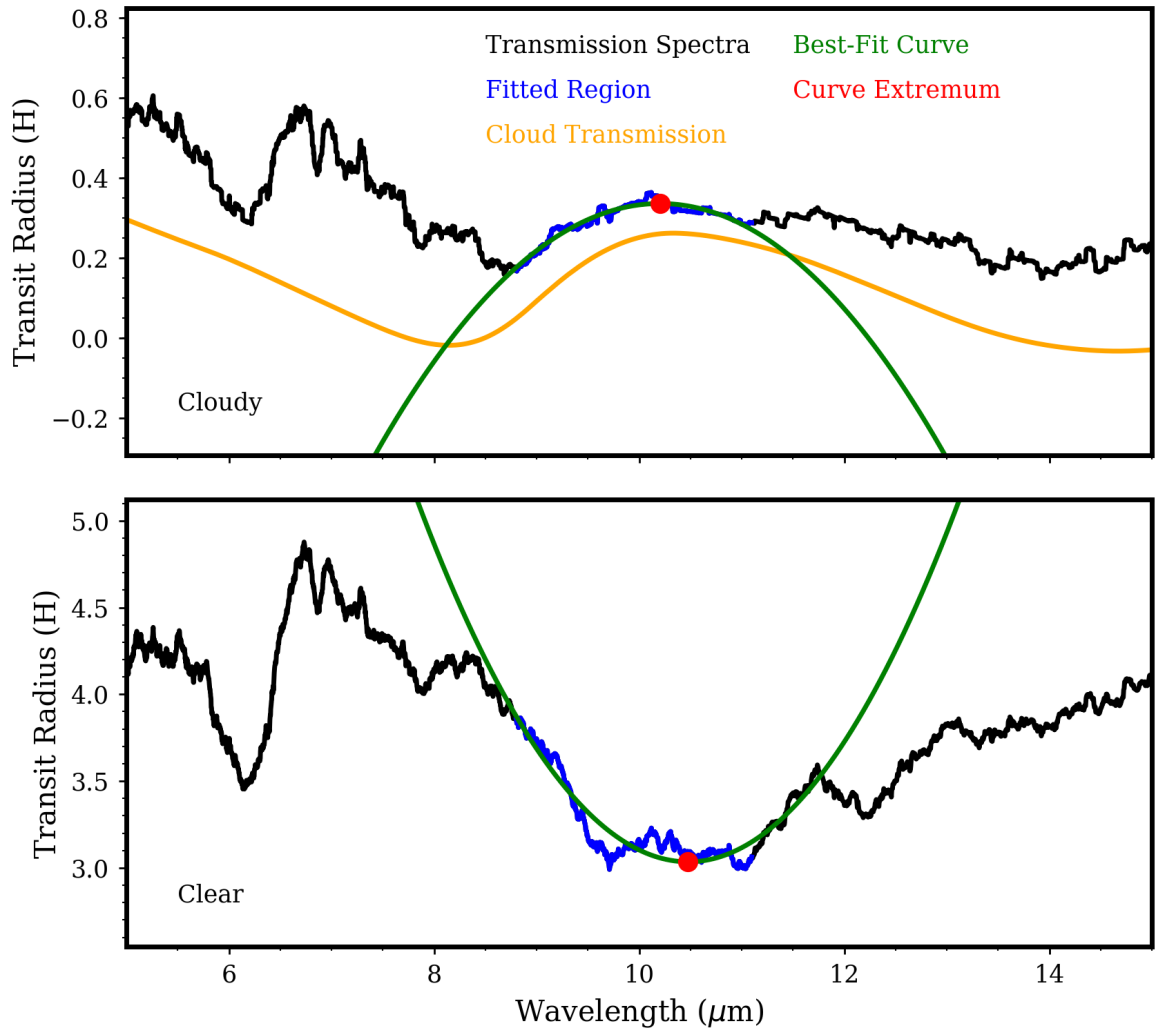

**Supplementary Figure 7. Obtaining the amplitude of the 10 μm silicate spectral feature.** We fit the model transmission spectra with a 2nd order polynomial between 8.8 and 11.1 μm to obtain the concavity therein. The amplitude of the silicate feature is then computed as the difference between the maximum and minimum of the polynomial within 8.8 and 11.1 μm.

**Supplementary Table 1.** Properties of exoplanet sample: Observed $A_H$ values and computed equilibrium temperatures from ref. 1, computed log of 1 bar gravities, and observed host star type. Errors on the observations and calculated values represent 1 standard deviation uncertainties.

| Planet | $T_{eq}$ (K) | $A_H$ (H) | log(g) (cm s$^{-2}$) | Reference* | Stellar Type+ |
|---|---|---|---|---|---|
| HAT-P-1 b | 1320±103 | 1.27±0.35 | 2.893±0.0201 | 7 | G0V |
| HAT-P-3 b | 1127±68 | 0.52±0.74 | 3.280±0.0979 | 8 | K |
| HAT-P-12 b | 958±28 | 0.42±0.25 | 2.774±0.0335 | 9 | K4 |
| HAT-P-17 b | 780±34 | 0.27±0.78 | 3.135±0.0558 | 8 | K |
| HAT-P-18 b | 843±35 | 0.51±0.28 | 2.712±0.0537 | 10 | K2 |
| HAT-P-26 b | 980±56 | 1.92±0.31 | 2.660±0.1358 | 8 | K1 |
| HAT-P-32 b | 1784±58 | 1.30±0.28 | 2.847±0.1300 | 8 | --- |
| HAT-P-38 b | 1080±78 | 2.03±0.66 | 3.007±0.0878 | 11 | G5 |
| HAT-P-41 b | 1937±74 | 1.96±0.45 | 2.866±0.3047 | 8 | --- |
| HD 149026 b | 1627±83 | 1.09±0.56 | 3.255±0.0725 | 8 | G0 |
| HD 189733 b | 1201±51 | 1.45±0.47 | 3.361±0.0317 | 8 | K2V |
| HD 209458 b | 1449±36 | 0.78±0.17 | 2.991±0.0269 | 8 | F8 |
| WASP-12 b | 2580±146 | 1.62±0.31 | 3.011±0.1484 | 8 | --- |
| WASP-17 b | 1632±126 | 0.44±0.35 | 2.762±0.1698 | 8 | F4 |
| WASP-19 b | 2037±156 | 1.60±0.58 | 3.155±0.0292 | 12 | G8V |
| WASP-29 b | 963±69 | 0.12±0.49 | 3.002±0.1231 | 8 | --- |
| WASP-31 b | 1576±58 | 1.14±0.42 | 2.713±0.0385 | 13 | --- |
| WASP-39 b | 1119±57 | 1.22±0.16 | 2.653±0.0540 | 14 | --- |
| WASP-43 b | 1374±147 | 0.95±0.46 | 3.727±0.0786 | 15 | K7V |
| WASP-52 b | 1300±115 | 1.33±0.28 | 2.869±0.0279 | 16 | K2V |
| WASP-63 b | 1508±69 | 0.39±0.30 | 2.683±0.1364 | 8 | --- |
| WASP-67 b | 1026±59 | 0.86±0.70 | 2.926±0.1231 | 8 | --- |
| WASP-69 b | 964±38 | 0.65±0.13 | 2.785±0.0548 | 8 | K5 |
| WASP-74 b | 1915±116 | 0.97±0.45 | 3.004±0.0965 | 8 | F9 |
| WASP-76 b | 2206±95 | 1.62±0.21 | 2.852±0.0276 | 17 | F7 |
| WASP-80 b | 824±58 | 0.51±0.19 | 3.145±0.0390 | 18 | K7V |
| WASP-101 b | 1552±81 | 0.13±0.27 | 2.810±0.0873 | 8 | F6 |
| WASP-121 b | 2358±122 | 2.31±0.41 | 2.945±0.0309 | 19 | F6V |
| XO-1 b | 1196±60 | 3.33±0.76 | 3.219±0.0864 | 8 | G1V |

*For planet mass and radius measurements used to calculate log(g), retrieved from the NASA Exoplanet Archive.

+Retrieved from the NASA Exoplanet Archive.

**Supplementary Table 2.** Base-10 log of saturation vapor pressure in units of bars and surface energies in units of ergs cm$^{-2}$ of the considered condensates. Temperature T is in units of K; [Fe/H] refers to the base-10 log of metallicity; and $p_a$ is the local atmospheric pressure in bars. The saturation vapor pressures of KCl, ZnS, Na$_2$S, MnS, and Cr are taken from ref. 20; that of Mg$_2$SiO$_4$ and Fe are from ref. 21; that of TiO$_2$ is from ref. 22; and that of Al$_2$O$_3$ is from ref. 23. The surface tension of molten KCl is taken from ref. 24 as a proxy for the surface energy of KCl cloud particles. The surface energy of ZnS is taken from ref. 25 assuming its sphalerite form. The surface energies of Na$_2$S and MnS are estimated following ref. 26, though we note that these two species lack experimental measurements of their surface energies. The surface energies of Cr and Fe are expressed using the Eötvös rule and appropriate material constants[27]. The surface energies for Mg$_2$SiO$_4$ and Al$_2$O$_3$ are taken from measurements of their melts[28]. The surface energy of TiO$_2$ is computed[29] from Gibbs free energy arguments.

| Species | log(P$_{sat}$) (bars) | Surface Energy (ergs cm$^{-2}$) |
|---|---|---|
| KCl | 7.611 - 11382/T | 160.4 - 0.07(T-273.15) |
| ZnS | 12.812 - 15873/T - [Fe/H] | 860 |
| Na$_2$S | 8.550 - 13889/T - 0.50[Fe/H] | 1033 |
| MnS | 11.532 - 23810/T - [Fe/H] | 2326 |
| Cr | 7.490 - 20592/T | 1642 - 0.2(T-2133.15) |
| Mg$_2$SiO$_4$ | 4.88 - 32488/T - 1.4[Fe/H] - 0.2log($p_a$) | 436 |
| Fe | 7.23 - 20995/T | 1862 - 0.39(T-1803.15) |
| TiO$_2$ | 9.5489 - 32457/T | 535.124 - 0.04396T |
| Al$_2$O$_3$ | 17.7 - 45893/T - 1.66[Fe/H] | 690 |

**Supplementary Table 3.** Mass density at 25 °C[30], latent heat, limiting vapor species and their collision diameter, and contact angle (calculated from Eq. 2 in the Methods assuming $\sigma_{xC} = 0$) of considered condensates. The contact angle for ZnS is that over KCl, while for all other condensates it is that over $TiO_2$. $TiO_2$ and KCl do not possess associated contact angles since they act as CCN. The latent heat of vaporization is calculated from the saturation vapor pressures and the Clausius-Clapeyron equation. The limiting vapor species for each condensate is the condensate themselves for phase transition species, and the reactant with lower mixing ratio for species that condense via chemical reaction. The collision diameter of the limiting vapor species for each condensate is used to calculate their molecular diffusion coefficients via Chapman-Enskog theory. The molecular diffusion coefficients are important in calculating the rates of condensation and evaporation[31]. The collision diameters of the atomic species are assumed to equal their isolated van der Waals diameter[32]. For KCl we multiply the hard sphere diameter[33] by 1.2 to approximate an "isolated" KCl molecule, in line with the finding[32] that isolated van der Waals diameters were 10-30% larger than crystallographic van der Waals diameters for the same species. For $TiO_2$ we convert the published monomer volume[22] into a collision diameter assuming a spherical volume. The collision diameter used in Chapman-Enskog theory is the average of the condensate collision diameter and that of the background gas, which we assume to be $H_2$/He in the primordial number density ratio 85.6:14.4[34]. Given a collision diameter of 2.89 Å for $H_2$[35] and 2.6 Å for He[36], we average them according to the primordial ratio to arrive at a mean background gas collision diameter of 2.85 Å.

| Species | Limiting Species | Mass density (g cm$^{-3}$) | Latent heat ($10^{10}$ ergs g$^{-1}$) | Collision diameter (Å) | Contact angle (°)+ |
|---|---|---|---|---|---|
| KCl | KCl | 1.988 | 2.923 | 3.08 | --- |
| ZnS | Zn | 4.04 | 3.118 | 3.665 | 81.7 |
| Na$_2$S | Na | 1.856 | 11.57 | 4.195 | 61 |
| MnS | Mn | 4.0 | 8.297 | 3.675 | 77.6 |
| Cr | Cr | 7.15 | 7.582 | 3.655 | 74.8 |
| Mg$_2$SiO$_4$ | Mg | 3.21 | 25.59 | 3.845 | 0.1* |
| Fe | Fe | 7.87 | 7.197 | 3.695 | 77.2 |
| TiO$_2$ | TiO$_2$ | 4.25 | 7.78 | 3.385 | --- |
| Al₂O₃ | Al | 3.99 | 32.56 | 3.825 | 43.6 |

*Imposed minimum.
⁺At 800 K.

**Supplementary Table 4.** Lower boundary fixed mixing ratios of condensates for a solar metallicity atmosphere, corresponding to elemental abundances taken from ref. 34. The mixing ratios are multiplied by 10 for the 10 x solar metallicity cases. $TiO_2$ and KCl abundances vary over many orders of magnitude with temperature and pressure and so their bottom boundary mixing ratios are different for different equilibrium temperatures, gravities, and metallicities (see Supplementary Figure 4).

| Species | Mixing Ratio (ppm) |
|---|---|
| KCl | Varies |
| ZnS | 0.076 |
| Na₂S | 3.34 |
| MnS | 0.541 |
| Cr | 0.887 |
| Mg₂SiO₄ | 59.36 |
| Fe | 57.8 |
| TiO₂ | Varies |
| Al₂O₃ | 4.937 |